# Guidelines to the Problem of Location Management and Database Architecture for the Next Generation Mobile Networks

Md. Mamun Ali Sarker, Md. Ashraf Hossain Khan, M. M. A. Hashem
*Department of Computer Science and Engineering*
*Khulna University of Engineering and Technology (KUET), Khulna-920300, Bangladesh*
mamunkuet@yahoo.com, lunik_cse@yahoo.com, mma_hashem@hotmail.com

**Abstract**

*In near future, anticipated large number of mobile users may introduce very large centralized databases and increase end-to-end delays in location registration and call delivery on HLR-VLR database and will become infeasible. After observing several problems we propose some guidelines. Multitree distributed database, high throughput index structure, memory oriented database organization are used. Location management guidelines for moving user in overlapping network, neighbor discovery protocol (NDP), and global roaming rule are adopted. Analytic model and examples are presented to evaluate the efficiency of proposed guidelines.*

## 1. Introduction

The next-generation mobile network will be an integrated global system that will support terminal mobility, personal mobility, service provider portability [1]. To support global roaming, a location-independent personal telecommunication number (PTN) for each user is desirable. Location management [2] focuses mainly on location registration and call delivery. Location tracking is based on two types of databases: the home location register (HLR), and the visitor location register (VLR) is shown in Figure 1 (a).

We observed several problems such as, how registration and call delivery procedure will occur in the overlapping network coverage [3]. While a mobile terminal (MT) is roaming the global world, for any update, the roamed network sends control and status message to the original HLR, which is heavy cost and creates network congestion. Mobile switching center (MSC) i.e. an MSC /VLR may not know the address of a MT's HLR, then the global title translation (GTT) is required. Due to expected much higher user density in future mobile networks, the updating and querying loads on the two-level HLR-VLR location databases will become infeasible [4]. Database index structure, such as AVL tree, B+ tree, linear hashing is costly [6]. The future database architecture will be distributed, but there is no standard rule for global roaming. In this paper, we propose some guidelines regarding to the above problems.

The rest of this paper is organized as follows: Section 2 describes the current location management of global system for mobile communications (GSM) and IS-41 architecture. Section 3 describes multitree database architecture and its location management. Section 4 proposes two efficient database indices. Section 5 describes organization of location databases, while analytical model and conclusion are presented in section 6 and 7, respectively.

## 2. Location management in GSM and IS-41

Location management procedures in personal communications systems (PCS) involve numerous operations in various databases. These databases record the relevant information of a mobile user, trace the user's location by updating the relevant database entries, and map the user's PTN to its current location. Each subscriber has a service profile stored in the HLR. The VLR, where a MT resides also keep's a copy of the user profile and is collocated with MSC, which controls a group of registration areas (RAs). Whenever an MT changes its RA, the HLR is updated to point to the new location, and the MT is deregistered from the old VLR. As an incoming call arrives, the called MT's HLR is queried to get the location of the serving VLR of the MT, then a routing address request message is sent to the MSC/VLR. The MSC allocates a temporary local directory number(TLDN) to the called MT and sends back TLDN to the HLR, which in turn relays this information to the calling MSC. A connection to the called MSC then can be set up through the SS7 signaling network.

As the number of subscribers increasing, the volume of signaling traffic generated is extremely high [4], [5]. New database architectures are required. Most of these are based on distributed database. We have considered distributed database architecture for better performance than two-level centralized database architecture.

## 3. Multi-tree distributed database architecture

### 3.1. A multi-tree structure for PCS location databases

Three-level tree database architecture [9] is illustrated in Figure 1(b), consisting of a number of distributed database subsystems (DSs), each of which is a three-level tree structure. These DSs communicate with each other only through their root databases, DB0s, which are connected to the others by the public switched telephone network (PSTN), ATM networks, or other networks. The databases "DB0" and "DB2" correspond to the HLR and the VLR in GSM, respectively. A number of DB2s are grouped into one DB1 and several DB1s are connected to a single DB0. All users' service profiles are stored in the DB0. The DB2 has a copy of the service profiles of the users currently roaming within its area.

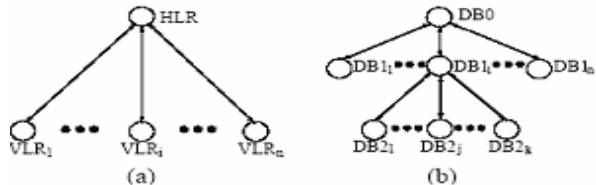

**Figure 1: (a) Two-level database architecture in GSM and (b) Three-level hierarchical database architecture in PCS.**

The multitree database architecture is motivated by the following:

It is much more robust than the one-root hierarchical architecture. The crash of one root database will not disrupt the operation of other root databases, and the recovery is much easier than in centralized database.

It is scalable and crucial to support continuously increasing number of mobile subscriber's in future mobile networks. When the capacity of a root database is saturated, a new DS is readily added. A location-independent PTN provides a basis for global roaming. A subscriber can retain its lifelong PTN regardless of its location and service provider.

It is easy to expand and maintain in the multioperator environment of a global mobile system. Each service provider can have its own DSs and it is straight-forward for a service provider to expand its service coverage by adding new DSs.

### 3.2. Neighbor discovery protocol

The concept of boundary interworking unit (BIU), boundary location register (BLR) and boundary location area (BLA) are same as [3] but works differently. We replaced the term BLR with neighbor location register (NLR) for its function and add extra facility for them.

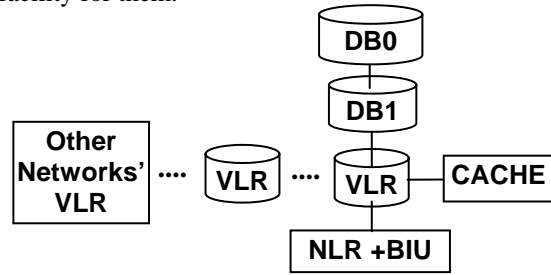

**Figure 2: Neighbor discovery protocol layout.**

The protocol works in several steps:

1. BIU finds own and other operator's network, VLR, signal strength etc. surrounding its VLR coverage. Store them to NLR.

2. BIU can also capture the MT at BLA which is under other network's coverage may be affected by them. Store them to NLR.

3. Recent registered, called or calling MTs with respective VLR, operator, connection networks are stored in cache temporarily.

4. VLR first searches from cache. If found, it establishes connection to that VLR without disturbing upper DB1 and DB0.

5. When any MT is affected by another overlapping network, the NLR supports the MT for correct registration and call delivery. The procedure is described later.

### 3.3. Location management for overlapping network coverage

Various situations from our own ideas are presented in figure for discussing registration and call delivery procedure for overlapping network coverage. Assume that, NLR and BIU are congested with each VLR. Here Network X is the own network of a discussing MT and Network Y, Z are roaming network. Procedures:

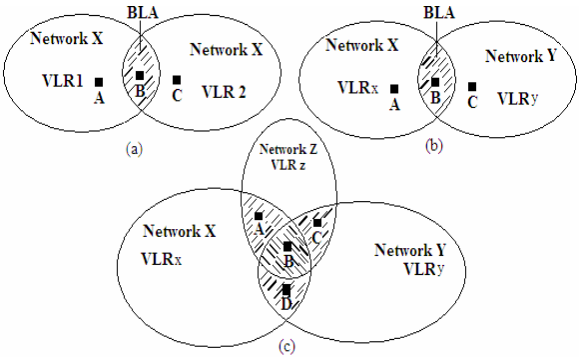

**Figure 3: location management for overlapping network, (a) for same network, (b) two diff network (c) several diff networks.**

(a) For same network:
1. For point A and C, MT is registered to VLR1 and VLR2 respectively, and no problem.
2. When MT is moving to point B, VLR2 is stimulated to register the MT. But MT is already in VLR1. MT requests to base station (BS) of VLR2 for bandwidth availability. NLR1 stores MT as overlapping zone adjacent to Y (from neighbor discovery protocol).
3. BS of VLR2 broadcasts their bandwidth range. MT calculates quality of service (QoS) q=(bandwidth required/bandwidth available) and velocity sign,
$V_s$ = (MT velocity with direction) / (fixed velocity)
for each adjacent VLR. Where velocity is positive for own network. Absolute of update condition C = [q.$V_s$] is determined. The lowest value of C is used for registering with that network.
4. Only if the MT is making call and moving toward another network, BIU makes an address pointer from NLR1 to NLR2 for further step. If MT is detached from VLR1 then NLR2 stores MT and requests BIU to support radio connection and continue its call delivery.
5. When MT is fully under control of NLR2, it requests its VLR to send register request to DB1 and to delete info from previous VLR.

(b) For two different networks:
1. Everything is same as (a), but only register request is send to DB1 to DB0 for diff network.

(c) For several networks:
1. For point A, C, D everything is same as (a). For point B, address pointer is each to each network. And when MT is fully registered to one network, all those address pointer is removed.
2. Here also update request is send to DB1 to DB0 for several networks.

### 3.4. Global roaming rule

Guidelines for global roaming rule:
1. Corresponding network should have the ability to route calls to MT regardless of its attachment to the network.
2. The user should have the ability to access their personal services independent of their point of attachment.
3. There should be distinguished universal personal telecommunication (UPT) number for each user.
4. There should be the ability that allows MT to receive their personalized end-to-end services regardless of their current networks.
5. It is required to develop global roaming agreements between different countries, regions and service providers, and to increase available radio spectrum based on these international agreements.
6. The integration of all networks will be controlled by PSTN.

## 4. Two efficient database indices

A database usually consists of two parts: an index file and a data file. There are two classes of indices: the disk-oriented index, such as the B+ tree and the memory –resident index, such as the AVL-tree and the T-tree. While the disk-oriented indices are designed primarily to minimize the number of disk space, the memory-resident indices aim to reduce computation time while using as little memory as possible. For real-time applications, the memory-resident indices are preferred due to their much faster access times than the disk-resident indices [6]. The AVL- tree has poor storage utilization since each node stores only one data item while requiring two pointers and some other control information.

### 4.1. The T-tree

The T-tree [6], which evolved from the AVL-tree and the B-tree, is a binary tree in which each node called T node contains a number of data items, a parent pointer, a left-child pointer, a right-child pointer is shown in Figure 4. The T-tree is fast since it retains the intrinsic binary search nature of the AVL-tree. It contains a number of data items in each node similar to the B-tree, thus having good storage utilization. All T-nodes are sorted in increasing order of the search values. To find a value in the T-tree, a search algorithm for the T-tree is needed. According to [6], one efficient search algorithm for the T-tree can be described as follows: (i) Each search begins with the root node; (ii) If the search value is less than the minimum value of the node, then the left-child node is searched.

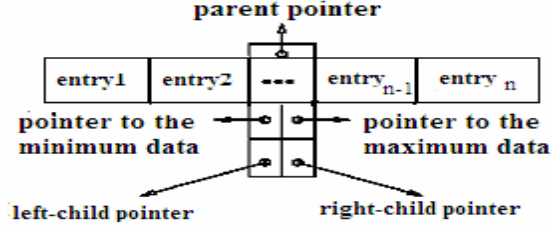

**Figure 4: T node of the T tree**

Otherwise, the current node is marked for future consideration and search goes down the subtree pointed to by the right-child pointer. When the search reaches a leaf, the last marked node is searched using a binary search. The search fails when the search value is not found in the marked node (bounding node) that bounds the search value or when the bounding node does not exist in the T-tree.

## 4.2. The direct file

In the direct file [7], there is a direct relationship between the record key and its storage location. One potential disadvantage of direct addressing is that space must be reserved for every possible key value, large storage space is wasted. However, when the number of possible key values is relatively close to the number of actual key values, direct addressing is very effective. The location-independent numbering plan makes direct addressing quite suitable for large centralized PCS databases.

## 5. Organizations of location databases in PCS

Organization of DB0: Each subscriber in the whole PCS system has an entry in the index file. If all user PTNs are allocated successively within the numbering plan, the T-tree or the B+-tree can not result in a smaller index file than the direct file, Other control information required in the T-tree or the B+-tree are not needed in the direct file. Thus, the direct file is the best choice for the centralized DB0. If the user is staying in another DS, the user's entry stores a reference pointing to the DB0 associated with that DS. The scalability feature is very useful for future PCS applications since the number of subscribers is expected to increase steadily.

Organization of DB1: Each DB1 only consists of one part – the index file, in which each user currently residing in the DB1 area has a data item. Each data item in the index file consists of two fields: the user's PTN and a pointer to the DB2 where the user is visiting. T-tree is preferable for the index file of the DB1.

Organization of DB2: Each of the lowest layer database DB2s consists of two parts: the index file and the data file. Each user currently residing in the DB2 area has an entry in the index. From the following section, we can see that the T-tree is preferable for the index file of the DB2.

## 6. Analytic model

Each database can be modeled as a $M/G/1$ queue. The service response time of $DB_i$, denoted by $T_i$, here $S_i$ is given in [8].

$$T_i = E[S_i] + \frac{\lambda_i(Var[S_i]+E^2[S_i])}{2(1-\lambda_i E[S_i])} \quad (1)$$

Now let $n_0$ be the number of DB2s in a DB0. $n_1$ be the number of DB2s in a DB1, $\lambda_c$ be the call rate originating from an RA, $\lambda_u$ be the location update request rate originating from an RA, $q_0$ be the probability that the user enters a different DB0 area, $q_1$ be the probability that the user enters a new DB1 area within the same DB0 area, $p_0$ be the probability that the caller and the callee are in different DB0 areas, $p_1$ be the probability that the caller and the callee are in the same DB0 area but different DB1 areas, and $p_2$ be the probability that the caller and the callee are in the same DB1 area but different DB2 areas. The arrival rates to the DB0, DB1, and DB2 can be derived as:

$\lambda_0 = n_0[(2q_0+q_1)\lambda_u + (2p_0+p_1)\lambda_c]$,
$\lambda_1 = n_1[(1+q_0+q_1)\lambda_u + (2p_0+2p_1+p_2)\lambda_c]$,
$\lambda_2 = 2\lambda_u + (1+p_0+p_1+p_2)\lambda_c$.

The end-to-end locations update delay:
$$T_u = 2T_2 + (1+q_0+q_1)T_1 + (2q_0+q_1)T_0 \quad (2)$$

Similarly, end-to-end call delivery delay:
$$T_d = (1+p_0+p_1+p_2)T_2 + (2p_0+2p_1+p_2)T_1 \\ + (2p_0+p_1)T_0 \quad (3)$$

Access cost of T- tree:
Service time of $DB_i$ is:
$$E[S_i] = \theta_s(Y_1,Y_2) + \delta(Y_1,Y_2)p_u \quad (4)$$
$$Var[S_i] = \delta^2(Y_1,Y_2)p_s p_u \quad (5)$$

Where $\delta(Y_1,Y_2)$ is derived in [9]

Selection of indices for location databases: Let $N_t$ be the total number of users in the whole PCS system, $\Phi_i$ be the available storage capacity for $DB_i$, $N_{i,e}$ be the number of user entries in the index file of $DB_i$, $E_i$ be the size of a user entry in the index, $N_t$ be the number of users currently residing in the $DB_i$ area, and $M$ be the size of a user service profile (for DB1, $M=0$). Note that for the DB0 $N_{0,e} = N_t \neq N_0$

1) The first choice – memory resident direct file:
$N_{i,e} = N_t$ for DB0, DB1, and DB2. To implement the direct file, we have: $N_t E_i + N_i M \leq \Phi_i$ (6)

For the DB0, if the searched MT is within the DB0 area, the service time of the DB0 is $S_0 = 2T_s$ (one access to the index file plus one access to the data file) where $T_s$ is the memory access time or the disk block access time; if the searched MT is not in this DB0 area, $S_0 = T_s$; if a service request comes from another DB0, $S_0 = 2T_s$. Thus, the expected service time of DB0,

$$E[S_0] = \left\{1 + \frac{(q_0+q_1)\lambda_u + (p_0+p_1)\lambda_c}{(2q_0+q_1)\lambda_u + (2p_0+p_1)\lambda_c}\right\}T_s \quad (7)$$

$$Var[S_0] = \frac{(q_0\lambda_u + p_0\lambda_c)[(q_0+q_1)\lambda_u + (p_0+p_1)\lambda_c]T_s^2}{[(2q_0+q_1)\lambda_u + (2p_0+p_1)\lambda_c]^2} \quad (8)$$

Applying (7) and (8) in (1), the response time for DB0, $T_0$ can be computed. For the DB1, the service time is fixed, i.e., $E[S_1] = T_s$ Thus, $Var[S_1] = 0$. For DB2, if the inquired MT is residing in the DB2 area, the service time is $S_2 = 2T_s$; otherwise, $S_2 = T_s$. The expectation and variance of DB2 is given by:

$$E[S_2] = \frac{4\lambda_u + (2+p_0+p_1+p_2)\lambda_c}{\lambda_2}T_s \quad (9)$$

$$Var[S_2] = \frac{(2\lambda_u + \lambda_c)(p_0+p_1+p_2)\lambda_c T_s^2}{\lambda_2^2} \quad (10)$$

2) The second choice – T-tree:
Let $a_1$ be the size of a node pointer and $a_2$ be the size of the pointer to the minimum (maximum) element in a node. Then, the size of a node is $3a_1 + 2a_2 + Y_1 E_i$. To implement the T-tree in the $DB_i$, we must have:

$$\frac{N_{i,e}(3a_1 + 2a_2 + Y_1 E_i)}{\kappa Y_1} + N_i M \leq \Phi_i \quad (11)$$

For the DB1 and DB2, $N_{i,e} = N_i$. The expectation and variance of the service time of $DB_i$ can be calculated using (4) and (5). The values of $p_s$ and $p_u$ can be computed for database $DB_i$. Given the values of $Y_1, Y_2, \kappa, c_1, c_2, c_3, c_4$, and $T_c$, we can compute the response time for a location database employing the T-tree as its index using expression (1).

3) The third choice – disk resident direct file:
Memory resident direct file is applicable to the disk-resident direct file except block access time, denoted by $T_b$, replaces, $T_s$, in all relevant formulas. The response time of $DB_i$ is obtaining using (1).

## 7. Simulation result

If $r_1$ % of the users has speed $\upsilon_1$, $r_2$ % has speed $\upsilon_2$, the average arrival rate of location updates generated in an RA is [8]: $\lambda_u = \frac{\rho L(\upsilon_1 r_1 + \upsilon_2 r_2)}{3600\pi}$ (12) where $\rho$ = user density, L= length of RA boundary. Call rate originating from an RA: $\lambda_c = \rho \xi A / 3600$ (13) where $\xi$ is the call origination rate per terminal /hr.

**Table 1: System parameter values [10], [11]**

| Parameter | $n_0$ | $n_1$ | $r_1$ | $r_2$ | $q_0$ |
|---|---|---|---|---|---|
| Value | 128 | 16 | 0.4 | 0.1 | 0.05 |
| Parameter | $q_1$ | L | $p_0$ | $p_1$ | $p_2$ |
| Value | 0.15 | 30.3 km | 0.01 | 0.04 | 0.45 |
| Parameter | A | $\upsilon_1$ | $\upsilon_2$ | $\xi$ | $\rho$ |
| Value | 57.4 $km^2$ | 5.6 km/hr | 56 km/hr | 1.4/hr / MT | 415 users/ $km^2$ |

Numerical example: We assume that $N_t = 1 \times 10^9$, $Y_1$ = 15, $\kappa$ = 0.95, $T_c = 1\mu s$, $c_1 = 10$, $c_2 = 100$, $c_3 = 20$, $T_s = 10\mu s$, $T_b = 20$ ms.

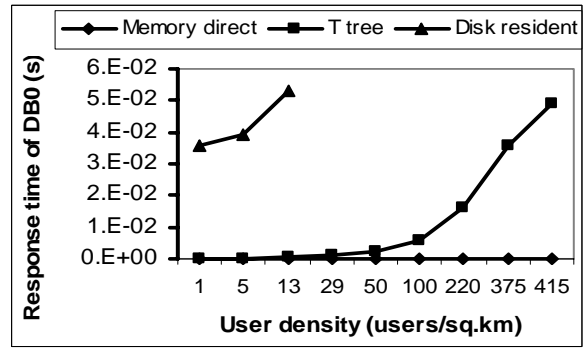

**Figure 5: The response time of DB0**

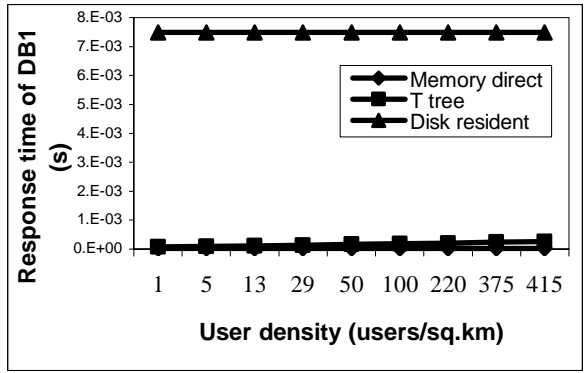

**Figure 6: The response time of DB1**

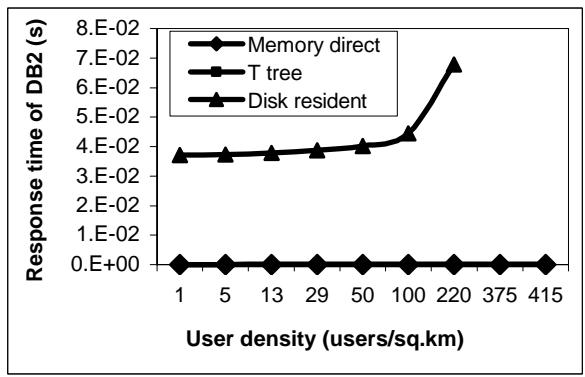

**Figure 7: The response time of DB2**

From Figure 5, 6, 7 we can see that the disk-resident indices incur much higher response time than memory-resident indices. As the user density approaches certain points, the disk-resident direct files are saturated. However, the response time of the T-tree and the memory-resident direct file increase very slowly. Based on the results shown in above figure, the memory-resident direct file is implemented in DB0.The T-tree is used both in DB1and DB2.

## 7. Conclusions

Analytic model and numeric result in figure reveal that incorporation of the three-level database architecture with the memory-resident direct file and the T- tree can achieve high throughput. It is scalable, robust, and efficient. Registration and call delivery in the overlapping network will reduce high paging and update signals between DB0(s). Network neighbor protocol is adopted for dynamic communication without disturbing corresponding DB0. Only distributed database architecture has also problems [3].

Our future work includes to reach a hybrid database architecture which lies between centralized and distributed database architecture.